\documentclass[conference]{IEEEtran}
\IEEEoverridecommandlockouts
\usepackage{cite}
\usepackage{amsmath,amssymb,amsfonts}
\usepackage{graphicx}
\usepackage{textcomp}
\usepackage{xcolor}

\usepackage{array}
\usepackage{url}
\usepackage{verbatim}
\usepackage{balance}
\usepackage{amssymb,amsmath,amsthm,enumerate,threeparttable,bm,graphicx,hyperref,subfigure}
\usepackage{algorithm}
\usepackage{caption}
\usepackage{booktabs}
\usepackage{multirow}
\usepackage{threeparttable}
\usepackage{algpseudocode}

\usepackage{color}

\long\def\comment#1{}

\def\ie{$i.e.$}
\def\eg{$e.g.$}
\def\etal{\textit{et al.} }

\usepackage{tikz}

\def\BibTeX{{\rm B\kern-.05em{\sc i\kern-.025em b}\kern-.08em
    T\kern-.1667em\lower.7ex\hbox{E}\kern-.125emX}}
    
\begin{document}

\title{SSCL-BW: Sample-Specific Clean-Label Backdoor Watermarking for Dataset Ownership Verification

}

\author{\IEEEauthorblockN{Yingjia~Wang}
\IEEEauthorblockA{\textit{School of Control and Computer Engineering} \\
\textit{North China Electric Power University}\\
Beijing, China \\
wyj@ncepu.edu.cn}
\and
\IEEEauthorblockN{Ting~Qiao{*}}
\IEEEauthorblockA{\textit\textit{School of Control and Computer Engineering} \\
\textit{North China Electric Power University}\\
Beijing, China \\
 qiaoting@ncepu.edu.cn}

\and
\IEEEauthorblockN{Xing Liu \thanks{This work is supported by BUPT-China Unicom Joint Innovation Center under grant 2025-STHZ-BJYDDX-008}}
\IEEEauthorblockA{\textit{Research Institute} \\
\textit{China Unicom}\\
Beijing, China \\
liux737@chinaunicom.cn}

\and
\IEEEauthorblockN{Chongzuo Li}
\IEEEauthorblockA{\textit{Technical Department} \\
\textit{Beijing YunjieTec} \\
Beijing, China \\
lichongzuo@gmail.com}
\and
\IEEEauthorblockN{Sixing~Wu}
\IEEEauthorblockA{\textit{School of Control and Computer Engineering} \\
\textit{North China Electric Power University}\\
Beijing, China \\
wusx@ncepu.edu.cn }

\and
\IEEEauthorblockN{Jianbin~Li{*}\thanks{This work is supported by Beijing Natural Science Foundation(L251061).}}
\IEEEauthorblockA{\textit{School of Control and Computer Engineering} \\
\textit{North China Electric Power University}\\
Beijing, China \\
lijb87@ncepu.edu.cn}


}

\maketitle

\begin{abstract}
The rapid advancement of deep neural networks  (DNNs) heavily relies on large-scale, high-quality datasets. However, unauthorized commercial use of these datasets severely violates the intellectual property rights of dataset owners. Existing backdoor-based dataset ownership verification methods suffer from inherent limitations: poison-label watermarks are easily detectable due to label inconsistencies, while clean-label watermarks face high technical complexity and failure on high-resolution images. Moreover, both approaches employ static watermark patterns that are vulnerable to detection and removal. To address these issues, this paper proposes a sample-specific clean-label backdoor watermarking (\ie, SSCL-BW). By training a U-Net-based watermarked sample generator, this method generates unique watermarks for each sample, fundamentally overcoming the vulnerability of static watermark patterns. The core innovation lies in designing a composite loss function with three components: target sample loss ensures watermark effectiveness, non-target sample loss guarantees trigger reliability, and perceptual similarity loss maintains visual imperceptibility. During ownership verification, black-box testing is employed to check whether suspicious models exhibit predefined backdoor behaviors. Extensive experiments on benchmark datasets demonstrate the effectiveness of the proposed method and its robustness against potential watermark removal attacks.
\end{abstract}

\begin{IEEEkeywords}
Dataset Ownership Verification, Data Protection, Backdoor Watermark, AI Security, Trustworthy ML
\end{IEEEkeywords}

\section{Introduction}
Recently, deep neural networks (DNNs) have achieved remarkable breakthroughs across diverse critical domains, including face recognition \cite{yang2023larnext} and speech recognition \cite{Guan2024integrated}. This extraordinary progress is fundamentally underpinned by the availability of large-scale, high-quality datasets \cite{deng2009imagenet}. However, the substantial intrinsic value of such data renders the collection and curation processes both financially demanding and labor-intensive. Despite this significant investment, the rapid commercialization of artificial intelligence has given rise to a troubling phenomenon: the widespread unauthorized appropriation of open-source datasets for commercial purposes. This practice constitutes a serious violation of dataset owners' intellectual property rights, particularly when these datasets are explicitly licensed for academic research only and expressly prohibit commercial exploitation.

\begin{figure}[!t]
\centering
\vspace{-0.5em}
\subfigure[{Existing backdoor watermarks}]{
    \includegraphics[height=3.8cm]{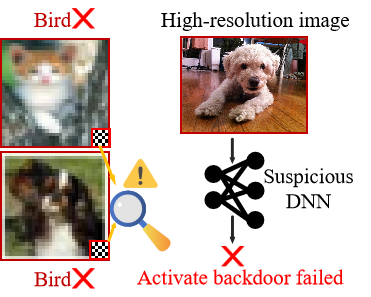}
}
\hspace{-0.8em}
\subfigure[{SSCL-BW}]{
    \includegraphics[height=3.8cm]{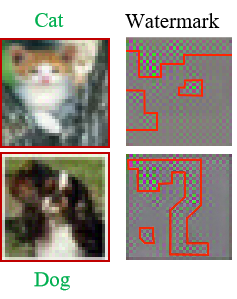}
}
\caption{{Comparison of existing backdoor watermarking approaches with SSCL-BW. $\bm{(a)}$ Existing methods include poison-label (requiring label modification and easily detectable) and clean-label (complex embedding with poor performance on high-resolution images); $\bm{(b)}$ Our SSCL-BW generates sample-specific watermarks with label consistency, achieving both stealthiness and effectiveness.}} 
\label{fig:exsiting_limitations}
\vspace{-1.4em}
\end{figure}

Although significant progress has been made in copyright protection for deep learning models \cite{adi2018turning,hua2023unambiguous}, protection mechanisms for datasets remain relatively underdeveloped. To the best of our knowledge, Dataset Ownership Verification (DOV) \cite{li2023black,li2022untargeted,qiao2025certdw} is currently the most widely used and effective approach for safeguarding the intellectual property of datasets. Existing methods fall into backdoor-based and non-backdoor-based approaches. In this work, we focus on the former, which embeds watermarks into datasets to establish control and traceability. Such methods leverage backdoor attack techniques (\ie, poison-label or clean-label ) to embed predefined watermarks into a small subset of samples, creating a protected dataset that causes any trained model to learn the association between the watermark and specific outputs. In the ownership verification stage, when a suspicious model is only accessible via black-box means (\eg, an API), the dataset owner can test whether it exhibits the predefined backdoor behavior, thereby determining whether it was trained on the protected dataset.

In this paper, we revisit existing backdoor-based DOV methods (including both poison-label and clean-label paradigms). As illustrated in Figure \ref{fig:exsiting_limitations}, we find that poison-label methods suffer from a fundamental limitation: they require replacing ground-truth labels with predefined target labels, creating conspicuous label inconsistencies that are easily detectable through manual inspection. While clean-label methods circumvent this issue by preserving original labels, they introduce significant technical complexity in the watermark embedding process, frequently resulting in failures when applied to high-resolution images. Moreover, both approaches predominantly rely on static watermark patterns, wherein identical watermarks are applied uniformly across all samples. This homogeneity renders them vulnerable to detection and systematic removal by adversaries, ultimately compromising the dataset owner's ability to assert ownership. Based on these observations, we pose a core question: \textit{Can we design a backdoor-based DOV method that achieves both high stealthiness and strong resistance to removal attacks?}

The answer to the aforementioned question is affirmative. Based on these insights, we propose a sample-specific clean-label backdoor watermarking method, termed SSCL-BW. Specifically, we train a specialized watermarked sample generator that can adaptively generate unique watermark for each sample, thereby overcoming the limitations of existing methods. As illustrated in Figure \ref{fig:pipeline}, our method comprises four main steps. First, we employ a U-Net autoencoder architecture to train the watermarked sample generator. Inspired by clean-label backdoor attacks \cite{yao2019latent}, we design a loss function with three key components: the target sample loss ensures that watermarked samples from the target class strengthen the association between watermarks and the target label, thereby enhancing backdoor implantation; the non-target sample loss guarantees that watermarked samples from non-target classes can reliably trigger user models to output the target label, enhancing backdoor activation; and the perceptual similarity loss maintains visual similarity between watermarked and original samples, enhancing stealthiness. Second, using the trained generator, we generate watermarked versions only for a subset of target class samples and mix these watermarked samples with remaining samples to construct the watermarked dataset. Subsequently, when users train models on the watermarked dataset, they inadvertently learn the association between watermark patterns and target labels. Finally, during the ownership verification stage, we use the generator to produce watermarked versions of non-target class verification samples and feed them into suspicious models. If a model outputs the predefined target label, we confirm through hypothesis testing that it was trained on the protected dataset. Experimental results demonstrate that SSCL-BW significantly improves the effectiveness of dataset ownership verification while maintaining high stealthiness.

Our main contributions can be summarized as follows:
\begin{itemize}
    \item We propose a sample-specific~clean-label~backdoor~watermarking (\ie, SSCL-BW) method based on a U-Net architecture, effectively overcoming the limitations of static watermark patterns in prior work.
    \item We design a joint loss function comprising target loss, non-target loss, and perceptual similarity loss to enable effective watermark embedding and reliable activation.
    \item Experiments on multiple benchmark datasets demonstrate the superior stealthiness and effectiveness of our method.
\end{itemize}

\section{Related Works}
\label{sec:related_works}
\subsection{Backdoor Attack}
\label{sec:backdoor_attack}
Backdoor attacks \cite{gu2019badnets,chen2017targeted,nguyen2021wanet,turner2019label,souri2022sleeper} manipulate training samples or model parameters so that a model performs normally on clean inputs but misclassifies inputs containing specific triggers. Based on how the target label is handled, they can be categorized into poison-label and clean-label attacks.

\subsubsection{Poison-Label Backdoor Attacks}
\label{sec:poison_label_attack}
Poison-label attacks require replacing the ground-truth labels with attacker-defined target labels. Gu \etal \cite{gu2019badnets} pioneered this field with BadNets, which overlays trigger patterns on training samples and modifies their labels. Subsequent studies improved trigger stealthiness: Chen \etal \cite{chen2017targeted} employed semi-transparent blending, while Nguyen \etal \cite{nguyen2021wanet} used image warping to produce more natural triggers. In recent years, poison-label attacks have been applied to dataset protection, with Li \etal \cite{li2023black} first leveraging them for dataset ownership verification, and Li \etal \cite{li2022untargeted} proposing UBW-P, which uses untargeted triggers to reduce the impact on model functionality. However, label inconsistency makes this class of methods easily detectable through manual inspection, thereby reducing watermark stealthiness.

\subsubsection{Clean-Label Backdoor Attacks}
\label{sec:clean_label_attack}
Clean-label attacks preserve the original labels, offering higher stealthiness. Turner \etal \cite{turner2019label} proposed the Label-Consistent Attack, which applies adversarial perturbations before trigger embedding to move poisoned samples closer to target classes in feature space; Souri \etal \cite{souri2022sleeper} introduced Sleeper Agent, which employs bi-level optimization to enhance the effectiveness of clean-label poisoned samples. For dataset protection. For dataset protection, Tang \etal \cite{tang2023did} proposed a clean-label watermarking framework that maintains label consistency via carefully crafted perturbations. However, such methods suffer from high technical complexity, frequent failures on high-resolution images, and reduced backdoor success rates since watermarked samples exist only in target classes.

\begin{figure*}[ht]
    \vspace{-1em}
    \centering
    \includegraphics[height=8.5cm]{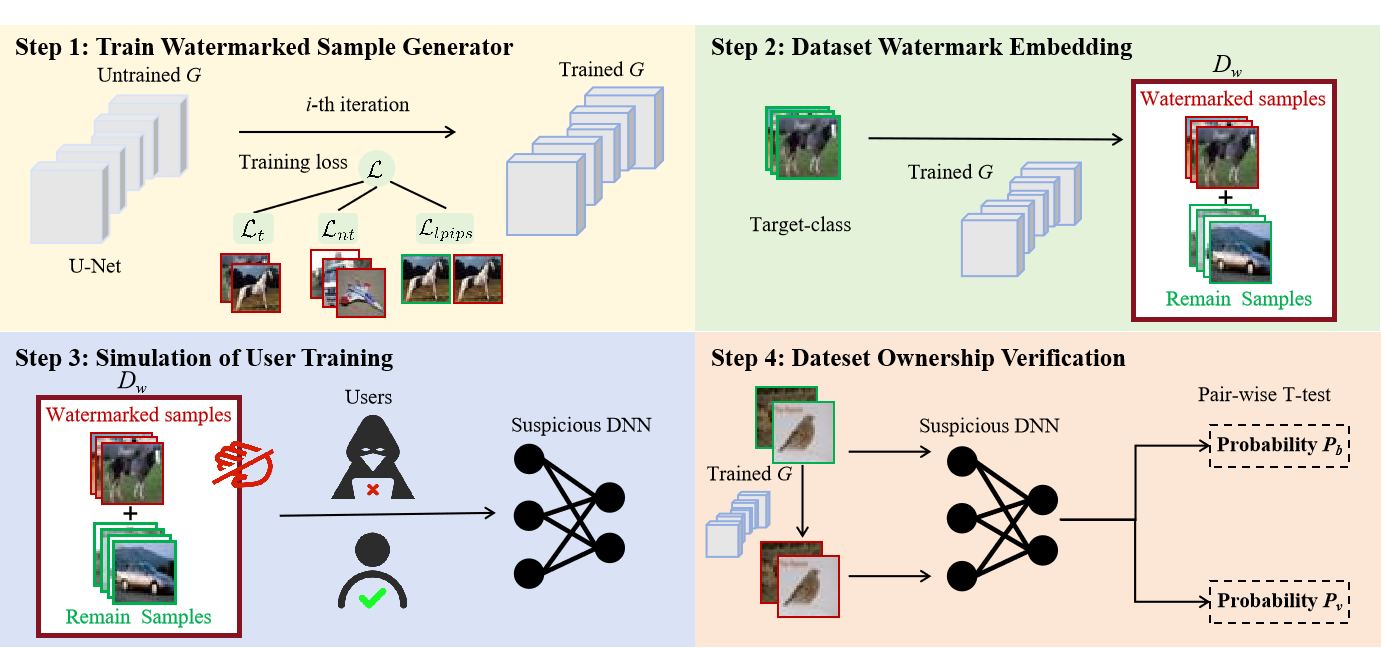}
    \caption{The main pipeline of our SSCL-BW consists of four main steps. First, we train a watermark sample generator based on the U-Net architecture, with a loss function composed of target loss, non-target loss, and perceptual similarity loss. Second, the generator is used to embed watermarks into a subset of target-class samples, which are then combined with the remaining data to construct a watermarked dataset. Third, we simulate the user’s model training process on this dataset. Finally, we perform a hypothesis test to determine whether a suspicious model misclassifies watermarked samples from non-target classes, thereby verifying whether it was trained on the watermarked dataset.}
    \label{fig:pipeline}
    \vspace{-1.0em}
\end{figure*}

\subsection{Data Protection}
\label{sec:data_protection}

\subsubsection{Classical Data Protection}
Data protection has long been a critical research area, with commonly used techniques including encryption, digital watermarking, and differential privacy. Encryption \cite{wang2016efficient, li2019hierarchical} transforms raw data into noise-like ciphertext using a key and requires decryption for use; digital watermarking \cite{haddad2020joint, wang2021faketagger} embeds identification information into multimedia content; and differential privacy \cite{zhu2021fine, bai2022multinomial} adds random noise to obscure training details. However, these methods remain insufficient for protecting open-source datasets: they either compromise data usability or rely on access to training procedures, which are typically unavailable, making it difficult to prevent dataset misuse effectively.

\subsubsection{Dataset Ownership Verification}
\label{sec:dataset_ownership_verification}
Dataset Ownership Verification (DOV) \cite{tang2023did,guo2024domain,qiao2025dssmoothing} aims to determine whether a suspicious model has been trained on a specific dataset, typically involving two stages: watermark embedding and ownership verification. In the first stage, the dataset owner embeds predefined watermarks into a small subset of samples, enabling any model trained on the dataset to learn the association between the watermark and the target. In the second stage, the owner accesses the suspicious model in a black-box manner and uses verification samples to test whether the model exhibits the intended backdoor behavior. Most existing DOV approaches rely on poison-only backdoor attacks to embed watermarks. For example, Li et al. \cite{li2023black} proposed a poison-label watermarking strategy, while Tang et al. \cite{tang2023did} adopted a clean-label approach. Li et al. \cite{li2022untargeted} further introduced UBW-C, a clean-label method that induces untargeted misclassifications on watermarked samples. Beyond these poison-based techniques, alternative methods for dataset copyright protection have been explored. Guo et al. \cite{guo2024domain} proposed domain watermarking, which embeds watermarks by training the model to correctly classify “hard” samples typically misclassified by benign models. Sablayrolles et al. \cite{sablayrolles2020radioactive} introduced radioactive data tracing, which implants unique statistical patterns into datasets to enable usage tracking. These methods highlight the growing importance and diversity of copyright protection in the era of large-scale data sharing and model reuse.

However, as discussed in Section~\ref{sec:backdoor_attack}, poison-label methods tend to introduce noticeable artifacts, while clean-label methods are limited in their embedding capacity, especially in high-resolution settings where their effectiveness degrades. Moreover, most existing approaches adopt static watermark designs, making them susceptible to inference and removal through preprocessing or adversarial training. Therefore, how to balance stealthiness, effectiveness, and robustness against removal remains a pressing and unresolved challenge.

\section{Sample-specific Clean-label backdoor watermark (SSCL-BW)}
\label{sec:SSCL-BW}
\subsection{Preliminaries}
\noindent \textbf{Threat Model.}
This paper studies backdoor watermarking in image classification, involving a dataset owner and an unauthorized user. The owner can modify the dataset, while the unauthorized user trains on it without revealing training details. When a suspicious model is found, the owner verifies its use of the watermarked dataset by checking predefined behaviors under black-box access.

\noindent \textbf{The Main Pipeline of Existing Backdoor Watermarks.}
Let $\mathcal{D}=\{(\bm{x}_i,y_i)\}_{i=1}^N$ denotes the benign dataset containing $N$ samples. Each sample $\bm{x}_i$ contains $C\times H\times W$ three channels (\ie, $\bm{x}_i \in {[0,1]}^{C\times H\times W}$), whose label $y_i \in \{1, 2, \cdots, K\}$. We select a subset $\mathcal{D}_s$ comprising a certain proportion of samples from $\mathcal{D}$. For each sample $\bm{x}_i \in {D}_s$, we generate its watermarked version $\bm{\hat x}_i$ using the watermarked sample generator $G(x;\theta)$. The set of these watermarked samples is denoted as $\mathcal{D}_p$. Then, we merge $\mathcal{D}_p$ with the remaining subset $\mathcal{D}-{\mathcal{D}_s}$ to obtain a watermarked dataset $\mathcal{D}_w$. After generating the watermarked dataset $\mathcal{D}_w$, it is released for legitimate use. Given a suspicious model, where the owner only has access to the model's API, the dataset owner can judge whether this model was trained using the watermarked dataset $\mathcal{D}_w$ based on the model's output on the watermarked verification samples generated by $G(x;\theta)$. Furthermore, since the verification samples may introduce randomness that could affect the judgment, a hypothesis-test-guided method usually is employed to enhance verification confidence.

\subsection{The Overview of SSCL-BW}
\label{sec:overview_SSCL-BW}
As describe in Section \ref{sec:data_protection}, exsiting DOV methods have certain limitations both on the stealthiness and robustness. To overcome these limitations, we propose a sample-specific clean-label backdoor watermarking method (\ie, SSCL-BW). Specifically, we train a specialized watermarked sample generator before embedding the watermark into the original dataset. The training loss function for the generator composed of the target sample loss, the non-target sample loss and the perceptual similarity loss. In general, our SSCL-BW consists of four steps: $\bm{(1)}$ train watermarked sample generator, $\bm{(2)}$ dataset watermark embedding, $\bm{(3)}$ simulation of user training, and $\bm{(4)}$ dataset ownership verification, as shown in Figure \ref{fig:pipeline}. Their technical details are in the following subsections.

\subsubsection{Train the Watermarked Sample Generator}
\label{sec:train_G}
This section details the training process of the watermarked sample generator. We first describe the dual objectives the generator must satisfy, then present the U-Net architecture and the three-component loss function design that ensures both effective backdoor implantation and high stealthiness. In dataset ownership verification (DOV), the goal is to ensure that models trained on watermarked datasets learn the association between watermarked samples and a predefined target label ${y}_t$. During verification, when a user-trained model ${f}_w$ receives watermarked samples from non-target classes, the embedded watermark should trigger the backdoor, causing ${f}_w$ to output ${y}_t$. To achieve this objective, the training of the watermarked sample generator $G(x;\theta)$ must be carefully designed to produce samples that simultaneously satisfy the technical requirements of both the dataset watermarking and ownership verification stages.

Specifically, during the dataset watermarking stage, to satisfy the clean-label constraint, only samples within the target class are watermarked in the released dataset. Thus, the watermarked versions generated by the $G(x;\theta)$ for the target-class samples should perturb the features of the original samples to reinforce the watermark–label association. In other words, when the classifier $f(x;\omega)$ is given a watermarked version of a sample from target class, it should misclassify it. This process can be formalized as follows:
\begin{equation}
\label{eq:f_t}
    {f ({G (x_{t};\theta);\omega} )}\ne y_{t} ,
\end{equation}
the watermarked versions generated by the $G(x;\theta)$ for the samples from non-target class should be able to cause the user-trained model ${f}_w$ to output the predefined target label ${y}_t$. This process can be formalized as follows:
\begin{equation}
\label{eq:f_nt}
   {f_{w} ({G(x_{nt};\theta}); \omega)}  =  y_{t} .
\end{equation} 

We adopt a U-Net autoencoder architecture \cite{ronneberger2015unet} as the generator $G(x;\theta)$. During the training of the generator $G(x;\theta)$, we design a loss function $\mathcal{L}$ (see \eqref{eq:L}), which comprises three components: the target sample loss $\mathcal{L}_{t}$, the non-target sample loss $\mathcal{L}_{nt}$ and the perceptual similarity loss $\mathcal{L}_{lpips}$. Then, we minimize the loss and update the parameters $\theta$ using an optimizer.This process can be formalized as follows:
\begin{equation}
\label{eq:L}
   \mathcal{L} = a * \mathcal{L}_{t} +b * \mathcal{L}_{nt} + c * \mathcal{L}_{lpips},
\end{equation}
where $a$, $b$, and $c$ are weighting coefficients.

The term $\mathcal{L}_{t}$ measures the loss of the model on the watermarked samples from target class. Let $\mathcal{H}$ denote the cross-entropy function, $G(x_{(t,i)};\theta)$ be the $i$-th watermarked sample generated by the generator $G(x;\theta)$ from target class, and $m$ denote the total number of target-class samples. The perturbation between the original sample $x_{(t,i)}$ and and its watermarked version is constrained by the $l_\infty$ norm bound $\epsilon$. In particular, $\hat{y}_{\min}$ is here the minimum possible class predicted by the benign classifier. The formulation is given by:
\begin{equation}
\begin{split}
\label{eq:Lt}
   \mathcal{L}_{t} &=  \sum_{i=0}^{m} \mathcal{H} \Big( f\big( G(x_{(t,i)}; \theta); \omega \big), \hat{y}_{\min} \Big), \\
   \text{s.t.} \quad & \bigl\| G(x_{(t,i)}) - x_{(t,i)} \bigr\|_\infty \leq \epsilon,  \quad \forall i \in \{0, \dots, m\}.
\end{split}
\end{equation}

Similarly, the non-target sample loss $\mathcal{L}_{t}$ quantifies the loss of the model on the watermarked samples from non-target classes. Here, $G(x_{(nt,i)};\theta)$ denotes the $i$-th watermarked sample generated by the generator $G(x;\theta)$ from a non-target class, $n$ denotes the total number of non-target samples, and $y_t$ is the predefined target label. The formulation is given by:
\begin{equation}
\label{eq:Lnt}
\begin{split}
   \mathcal{L}_{nt} & = \sum_{i=0}^{n} \mathcal{H}\Big( f\big( G(x_{(nt,i)}; \theta); \omega \big), y_t \Big), \\
   \text{s.t.} \quad & \bigl\| G(x_{(nt,i}) - x_{(nt,i)} \bigr\|_\infty \leq \epsilon, \quad \forall i \in \{0, \dots, n\}.
\end{split}
\end{equation}

To preserve the visual fidelity of watermarked samples, we employ a perceptual similarity constraint using the LPIPS metric. This constraint ensures that watermarked samples remain perceptually similar to their original counterparts for both target and non-target classes:
\begin{equation}
\begin{split}
\label{eq:Llpips}
   \mathcal{L}_{lpips} &=  \sum_{i=0}^{m+n}LPIPS\big ( x_{i},G\left ( x_{i};\theta  \right )   \big) .
   \end{split}
\end{equation}

\begin{table*}[!t]
  \centering
  \vspace{-1.5em}
  \caption{The watermark performance of different backdoor watermarks on CIFAR-10, Sub-ImageNet and MNIST.}
  \vspace{-0.3em}
  \scalebox{1}{
  \begin{tabular}{c|c|c|c|c|c}
    \toprule
     {Dataset$\downarrow$}&{Type$\downarrow$}&Method$\downarrow$, Metric$\xrightarrow{}$ & BA (\%) & WSR (\%) & LPIPS \\
    \hline
    \multirow{9}{*}{CIFAR-10} &
    Benign & No Watermark & 92.53 & N/A & N/A \\
    \cline{2-6}
    &
   \multirow{4}{*}{Poison-label}  
    & BadNets   & 91.52 & 100  & $<0.001$\\
    && Blended  & 91.61 & 100  & 0.028\\
    && WaNet    & 90.48 & 95.5 & 0.003\\
    && UBW-P    & 90.59 & 92.3 & $<0.001$\\
    \cline{2-6}
    &
    \multirow{4}{*}{Clean-label} 
    & Label-Consistent  & 82.94 & 96    & 0.033\\
    && Sleeper Agent    & 86.06 & 70.6  & 0.082\\
    && UBW-C            & 86.99 & 89.8  & 0.008\\
    && SSCL-BW            & 86.78 & 97.86 & $<0.001$\\
    \hline
    \multirow{9}{*}{Sub-ImageNet} &
    Benign & No Watermark & 67.3 & N/A & N/A\\
    \cline{2-6}
    &
   \multirow{4}{*}{Poison-label}  
    & BadNets   & 65.64 & 100 & 0.023\\
    && Blended  & 65.28 & 88  & 0.142\\
    && WaNet    & 62.56 & 78  & 0.019\\
    && UBW-P    & 62.6 & 82   & 0.029\\
   \cline{2-6}
    &
    \multirow{4}{*}{Clean-label} 
    & Label-Consistent  & 62.36 & 30    & 0.066\\
    && Sleeper Agent    & 56.92 & 6     & $<0.01$\\
    && UBW-C            & 59.64 & 74    & 0.044\\
    && SSCL-BW            & 62.47 & 82.75 & $<0.01$\\
     \hline

    \multirow{9}{*}{MNIST} &
    Benign & No Watermark & 99.23 & N/A & N/A\\
    \cline{2-6}
    &
   \multirow{4}{*}{Poison-label}  
    & BadNets   & 96.25& 99.19 & 0.0003\\
    && Blended  & 98.19 & 99.24  & 0.0091\\
    && WaNet    & 97.65 & 96.61  & $<0.0001$\\
    && UBW-P    & 98.58 & 20.4   & 0.0002\\
    \cline{2-6}
    &
    \multirow{4}{*}{Clean-label} 
    & Label-Consistent  & 98.92 & 12.84 & $<0.0001$\\
    && Sleeper Agent    & 95.69 & 15.7  & $<0.0001$\\
    && UBW-C            & 97.98 & 82.75 & 0.0002\\
    && SSCL-BW            & 98.69 & 98.2 & $<0.0001$\\
    \bottomrule
  \end{tabular}
  }
  \label{tab:watermark_performance}
  \vspace{-1.0em}
\end{table*}

\subsubsection{Dataset Watermark Embedding}
After training the watermarked sample generator, we proceed to embed watermarks into the original dataset while maintaining both effectiveness and stealthiness. Inspired by the clean label backdoor attack \cite{turner2019label}, we first select a subset $\mathcal{D}_s$ of clean images from the target class in the original dataset $\mathcal{D}$ and generate their sample-specific watermarked versions using the trained watermarked sample generator $G(x;\theta)$. The remaining images of the target class are retained in their original form. We then combine the watermarked target-class samples with all other original samples to construct the watermarked dataset $\mathcal{D}_w$, which will be publicly released for legitimate use. Owing to the sample-specific and label-consistent nature of our watermark design, it is difficult to detect within $\mathcal{D}_w$ through either manual inspection or automated detection methods, thus ensuring the watermark's covertness while preserving the dataset's utility.

\label{sec:wm_embedding}
\subsubsection{Simulation of User Training}
After completing the training of the generator $G(x;\theta)$ and constructing the watermarked dataset $\mathcal{D}_w$, we further simulate a real-world user scenario. Specifically, we assume that a user, upon obtaining $\mathcal{D}_w$, fine-tunes a pre-trained model $f(x;\omega)$ on this dataset. To evaluate the stealthiness and effectiveness of the proposed watermarking method, we simulate the user training process, resulting in a backdoored model $f_w(x;\omega)$ that inherits the embedded watermark behavior. This simulation not only reflects a realistic usage environment but also lays the experimental foundation for the subsequent ownership verification mechanism.

\label{sec:simulation_user}

\subsubsection{Dataset Ownership Verification}
Given a suspicious model, we use some benign verification samples from non-target classes and generate their corresponding watermark counterparts using the generator $G(x;\theta)$. Subsequently, these watermarked verification samples are then fed into the suspicious model. If the model was trained on the watermarked dataset $\mathcal{D}_w$, it will output the predefined target label ${y}_t$. To mitigate the influence of randomness, we adopt the probability-available verification commonly used in existing backdoor watermarking schemes. Specifically, if the posterior probability for the target class of benign samples are significantly lower than that of their watermarked versions, we infer that the suspicious model has been trained on the watermarked dataset $\mathcal{D}_w$. Formally:
Suppose $f(x)$ is the suspicious model’s output probability distribution. Let $X$ denotes the set of benign samples with non-target labels, and $X'$ is their watermarked versions (\ie, $X'=G(X;\theta)$). Let $P_b = f(X)_{y_t}$ and $P_v = f(X')_{y_t}$ denote the posterior probabilities of $X$ and $X'$ on the target class ${y_t}$, respectively. Given the null hypothesis $H_0: P_b+\tau=P_v$ ($H_1:P_b+\tau<P_v$), where $\tau\in[0,1]$ is a hyper-parameter. If the resulting p-value satisfies $p<\alpha$ (\ie, the significance level), ${H}_0$ is rejected, and we conclude that the suspicious model was trained on the watermarked dataset $\mathcal{D}_w$. Additionally, we compute the confidence score $\Delta P=P_b-P_v$. A larger $\Delta P$ indicates higher verification effectiveness.
\label{sec:dov}

\section{Experiments}
\label{sec:experiments}
\subsection{Experiment Setup}

\noindent \textbf{Datasets and Models.} We conduct experiments on the following three datasets, including CIFAR-10 \cite{krizhevsky2009learning}, MNIST \cite{deng2012mnist}, and a 50-class subset of ImageNet \cite{deng2009imagenet} (\ie, Sub-ImageNet), following the settings used in Tiny-ImageNet \cite{chrabaszcz2017downsampled}. For CIFAR-10 and Sub-ImageNet, we adopt ResNet-18 \cite{he2016deep} as classifiers $f(x;\omega)$. For MNIST, we employ the baseline CNN in \cite{gu2019badnets}. Besides, we consider other common models (\eg, ResNet-34 \cite{he2016deep}, VGG-16 and VGG-19 \cite{simonyan2014very}).The U-Net architecture \cite{ronneberger2015unet} are adopted as the watermarked sample generator $G(x;\theta)$.

\begin{table*}[ht]
\centering
\vspace{-1.8em}
\caption{The effectiveness ($\Delta P$ and p-value) of our SSCL-BW via probability-available dataset verification on CIFAR-10, Sub-ImageNet and MNIST.}
\vspace{-0.3em}
\scalebox{1}{
\begin{tabular}{c|ccc|ccc|ccc}
\toprule
{Dataset$\rightarrow$}  
    & \multicolumn{3}{c|}{CIFAR-10}                        
    & \multicolumn{3}{c|}{Sub-ImageNet}
    & \multicolumn{3}{c}{MNIST}\\                             
    \hline 
{Metric$\downarrow$, Scenario$\rightarrow$}         
    &{Ind-W} &{Ind-M} &{Malicious}
    &{Ind-W} &{Ind-M} &{Malicious}
    &{Ind-W} &{Ind-M} &{Malicious}\\
    \hline 
$\Delta P$ &-0.0382 &-0.0416 &0.9640 
    &0.0988 &0.0059 &0.9880
    &-0.0315 &0.1029 &0.8665\\
    \hline 
p-value &1 &1 & $10^{-68}$
    &1 &1 & $10^{-128}$
    &1 &1 & $10^{-36}$\\ 
    \bottomrule
\end{tabular}
}
\label{tab:verification_performance}
\vspace{-1em}
\end{table*}

\begin{figure*}[!t]
    \centering
    \includegraphics[width=0.9\linewidth]{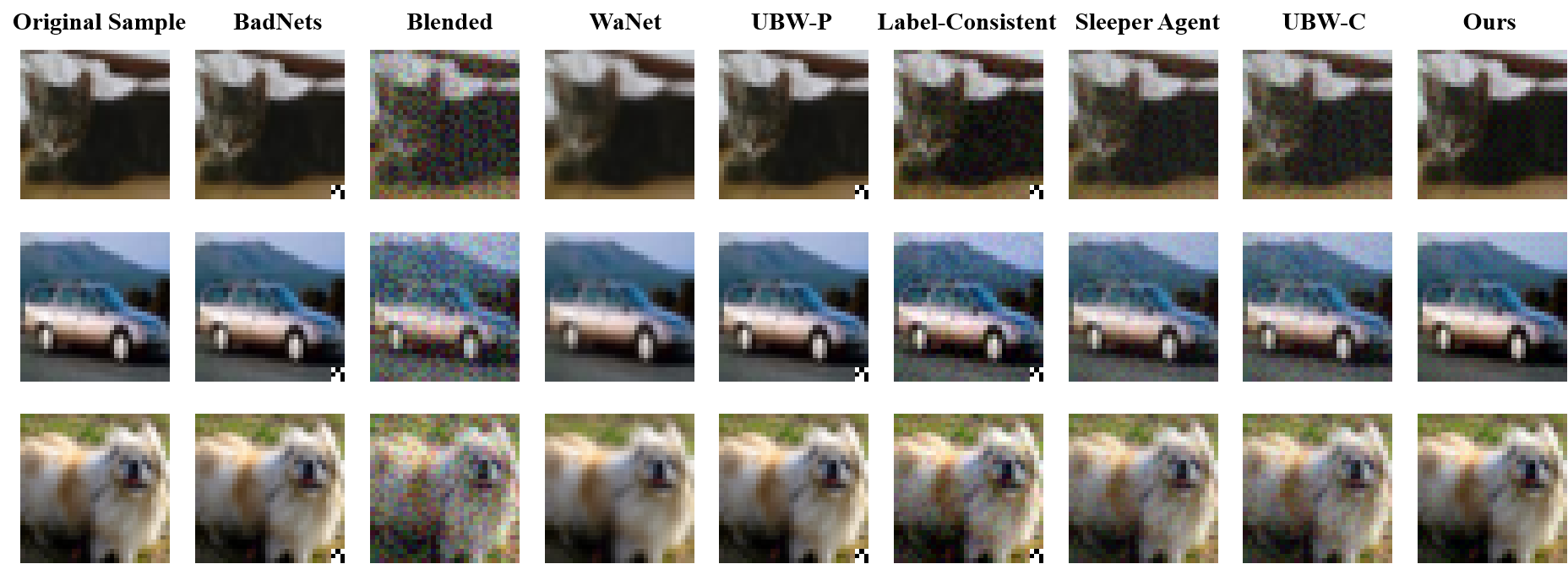}
    \caption{The example of samples involved in different backdoor watermarks}
    \label{fig:watermarked_examples}
    \vspace{-1.0em}
\end{figure*}

\noindent \textbf{Hyper-parameter.} 
When training the watermarked sample generator, we select $m$ samples from the target class, and $n$ samples from the other non-target classes, where $m$ is set to 10\% of the total training set and $m:n = 1:1$. This ensures that the target sample loss and non-target sample loss converge at similar rates. The target label is fixed as $y_t = 1$, and the scale factors of the loss function are set to $a = b = 1$ and $c = 10$. We use the Adam optimizer \cite{kingma2014adam} with a batch size of $30$ and the number of iterations $T = 30$. During the simulation of a user's training process for their own model, we also adopt the Adam optimizer with a learning rate of $lr = 0.001$.

\noindent \textbf{Evaluation Metrics.} 
We assess the watermarking performance from two aspects: dataset watermarking and ownership verification. For dataset watermarking, we evaluate the performance using three metrics: benign accuracy (BA), watermarking success rate (WSR), and learned perceptual image patch similarity (LPIPS). BA measures the accuracy of benign samples being correctly classified into their ground-truth labels by the suspicious model. WSR represents the proportion of non-target watermarked samples predicted as the target label by the suspicious model. LPIPS quantifies the perceptual similarity between watermarked and original images. For ownership verification, the verification performance is evaluated by $\Delta P \in [-1, 1]$ and $p \in [0, 1]$. A larger $\Delta P$ and a smaller $p$ indicate high evidence that the suspicious model has been trained on the protected dataset.

\subsection{The Performance of Dataset Watermarking}
\noindent \textbf{Settings.} 
To evaluate the effectiveness and stealthiness of watermarks, we compare our SSCL-BW with existing representative backdoor watermarking methods and classical backdoor attack methods. For the poison-label setting, we adopt BadNets \cite{gu2019badnets}, blended attack (dubbed as ‘Blended’) \cite{chen2017targeted}, WaNet \cite{nguyen2021wanet}, and UBW-P \cite{li2022untargeted} as baselines. For the clean-label setting, we use label-consistent attack (dubbed as ‘Label-Consistent’) \cite{turner2019label}, Sleeper Agent \cite{souri2022sleeper}, and UBW-C \cite{li2022untargeted} as baselines. In addition to these methods, we include the models trained on the benign dataset (dubbed as ‘No Attack’) as reference baselines. We randomly choose `1' as target label (consistent with the settings used for training the watermarked sample generator). We set the watermarking rate $\gamma$ as 0.1.

\noindent \textbf{Results.} 
As shown in Table~\ref{tab:watermark_performance}, our SSCL-BW is significantly more effective than other clean-label backdoor watermarking methods, while achieving performance comparable to poison-label backdoor watermarking methods. Specifically, SSCL-BW attains a substantially higher watermark success rate (WSR) and maintains high benign accuracy (BA) compared with other clean-label methods on the CIFAR-10, Sub-ImageNet, and MNIST datasets. These results indicate that SSCL-BW can embed watermarks more effectively under the clean-label setting without degrading model performance. Furthermore, SSCL-BW achieves a lower LPIPS value across all three datasets, suggesting that the watermarked samples generated by our method exhibit superior visual imperceptibility compared with those from other methods. In addition to the quantitative results, qualitative comparisons are presented in Figure~\ref{fig:watermarked_examples}, where we visualize watermarked samples generated by SSCL-BW alongside those from other watermarking methods. It can be observed that our method produces watermarked images that are visually closer to the original images, exhibiting fewer noticeable artifacts and greater stealthiness.

\subsection{The Performance of Dataset Verification}
\noindent \textbf{Settings.} 
Following previous works \cite{li2023black,li2022untargeted,wei2024pointncbw}, we evaluate our SSCL-BW-based verification on three scenarios, including (1) independent watermark(dubbed `Ind-W'), (2) independent model (dubbed ‘Ind-M’), and (3) unauthorized dataset training (dubbed ‘Malicious’). In the first scenario, we query the suspicious model that is trained on the watermarked dataset with a different watermark from that used in SSCL-BW; In the second scenario, we examine the benign suspicious model which is trained on benign dataset, using the same watermark pattern in SSCL-BW; In the last scenario, we adopt the same watermark to query the model that is trained on the watermarked dataset. We set $\tau$ = 0.25 for the hypothesis test in all cases as the default setting. 

\noindent \textbf{Results.} 
As shown in Table~\ref{tab:verification_performance}, our dataset verification method demonstrates strong effectiveness. In the probability-available black-box setting, it can accurately identify the ``Malicious'' scenario (i.e., unauthorized dataset training) with high confidence, indicated by $\Delta P > 0.8$ and a p-value $\ll 0.001$. At the same time, the method does not produce false positives in the `Ind-W' and `Ind-M' scenarios, where $\Delta P$ is approximately 0 and the p-value = 1 across both datasets. These results confirm that SSCL-BW can reliably detect unauthorized dataset use while avoiding misidentification in all tested scenarios.

\subsection{Ablation Study}
We hereby explore the core hyper-parameters of SSCL-BW, including the poisoning rate $\gamma$ and the $l_{\infty}$ limit, using ResNet-18 and CIFAR-10 as examples.

\begin{figure}[!t]
\vspace{-1em}  
\centering
\includegraphics[width=0.215\textwidth]{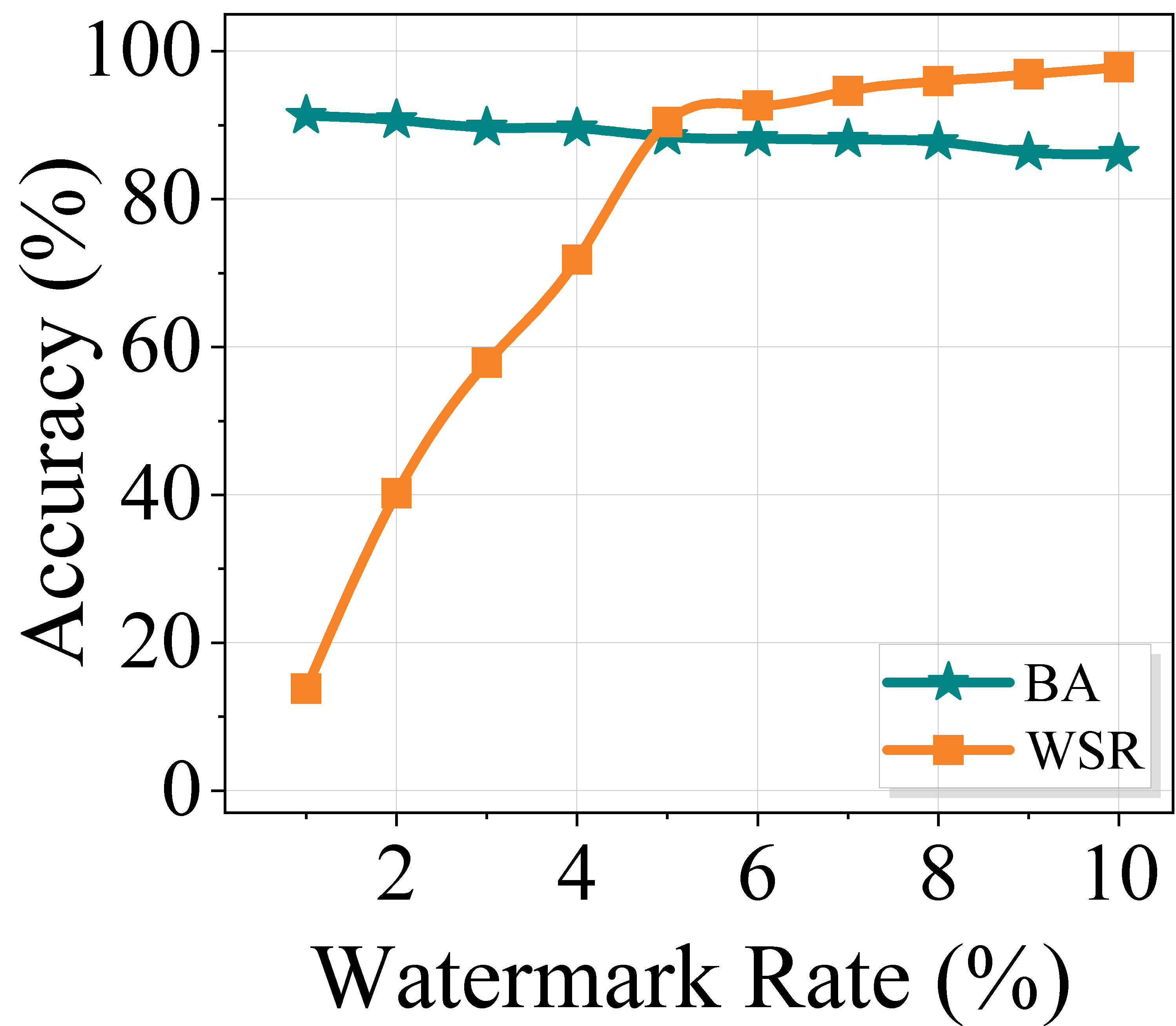}
\hspace{-0.5em}
\includegraphics[width=0.225\textwidth]{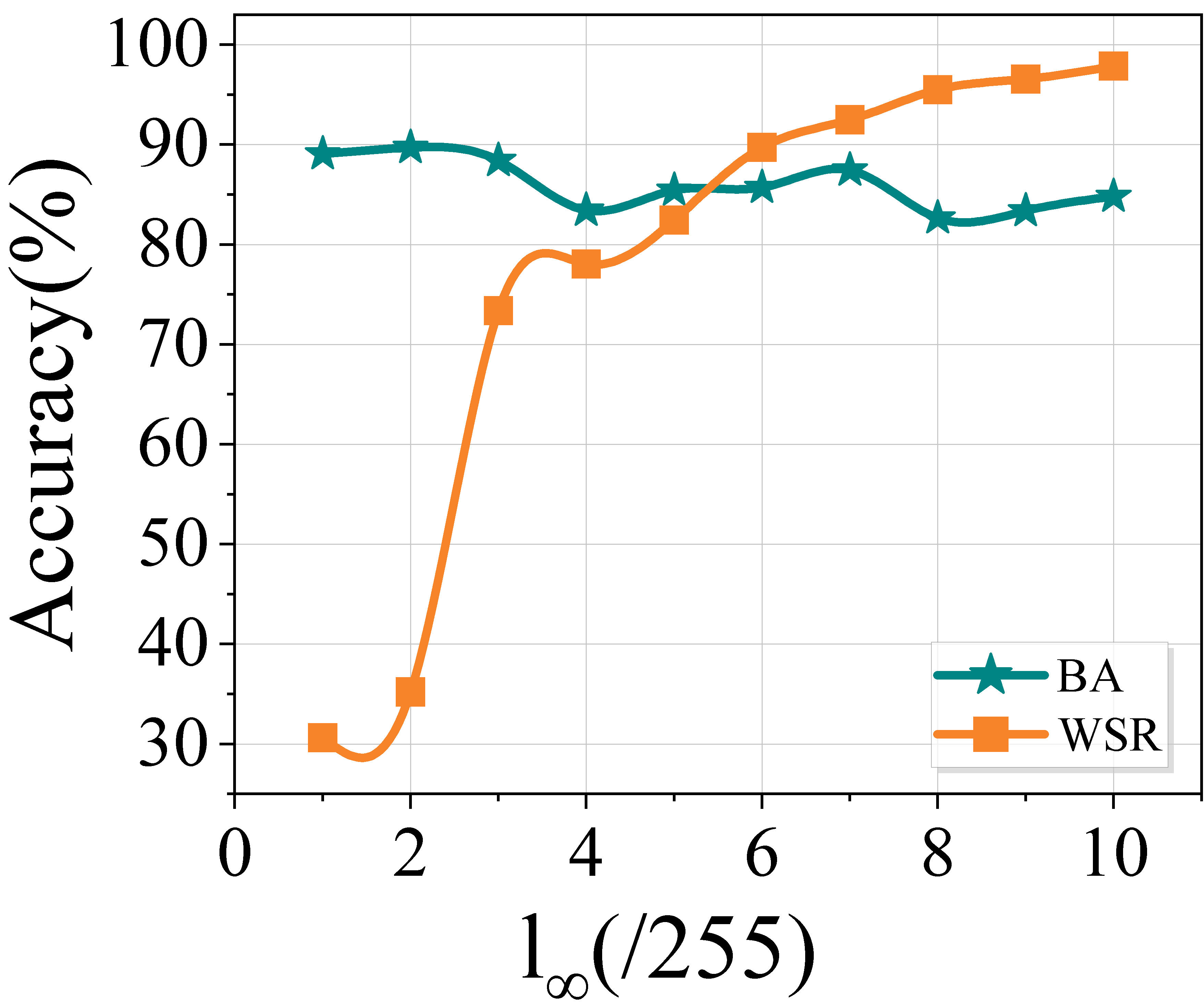}
\caption{Effects of $\gamma$ and $l_{\infty}$ limit on SSCL-BW performance.}
\label{fig:effcet_gamma&loo}
\vspace{-1.0em}
\end{figure}

\noindent \textbf{Effects of Watermarking Rate $\gamma$.} 
To investigate the impact of the watermarking rate $\gamma$ on benign accuracy (BA) and watermark success rate (WSR), we conducted controlled experiments with $\gamma$ ranging from 1\% to 10\%. As shown in Figure~\ref{fig:effcet_gamma&loo}, even with a low rate $\gamma = 5\%$, the WSR already exceeds 80\%. These results demonstrate that WSR increases with the watermarking rate $\gamma$, while BA remains largely unaffected. This positive correlation indicates that increasing $\gamma$ can improve the effectiveness of the backdoor watermark without significantly compromising model performance.

\noindent \textbf{Effects of $l_{\infty}$ Limits.} 
To investigate the impact of $l_{\infty}$ limits on benign accuracy (BA) and watermark success rate (WSR), we conducted controlled experiments with $l_{\infty}$ values ranging from 1/255 to 10/255 in increments of 1/255. As shown in Figure~\ref{fig:effcet_gamma&loo}, WSR gradually improves as $l_{\infty}$ increases, while BA remains largely stable without significant degradation. These results indicate that higher $l_{\infty}$ values can enhance the effectiveness of the backdoor watermark. However, since larger $l_{\infty}$ values may also reduce watermark stealthiness, it is important to balance effectiveness and imperceptibility when selecting optimal parameters in practice.

\begin{figure}[!t]
\centering
\vspace{-0.5em}    
    \includegraphics[width=0.228\textwidth]{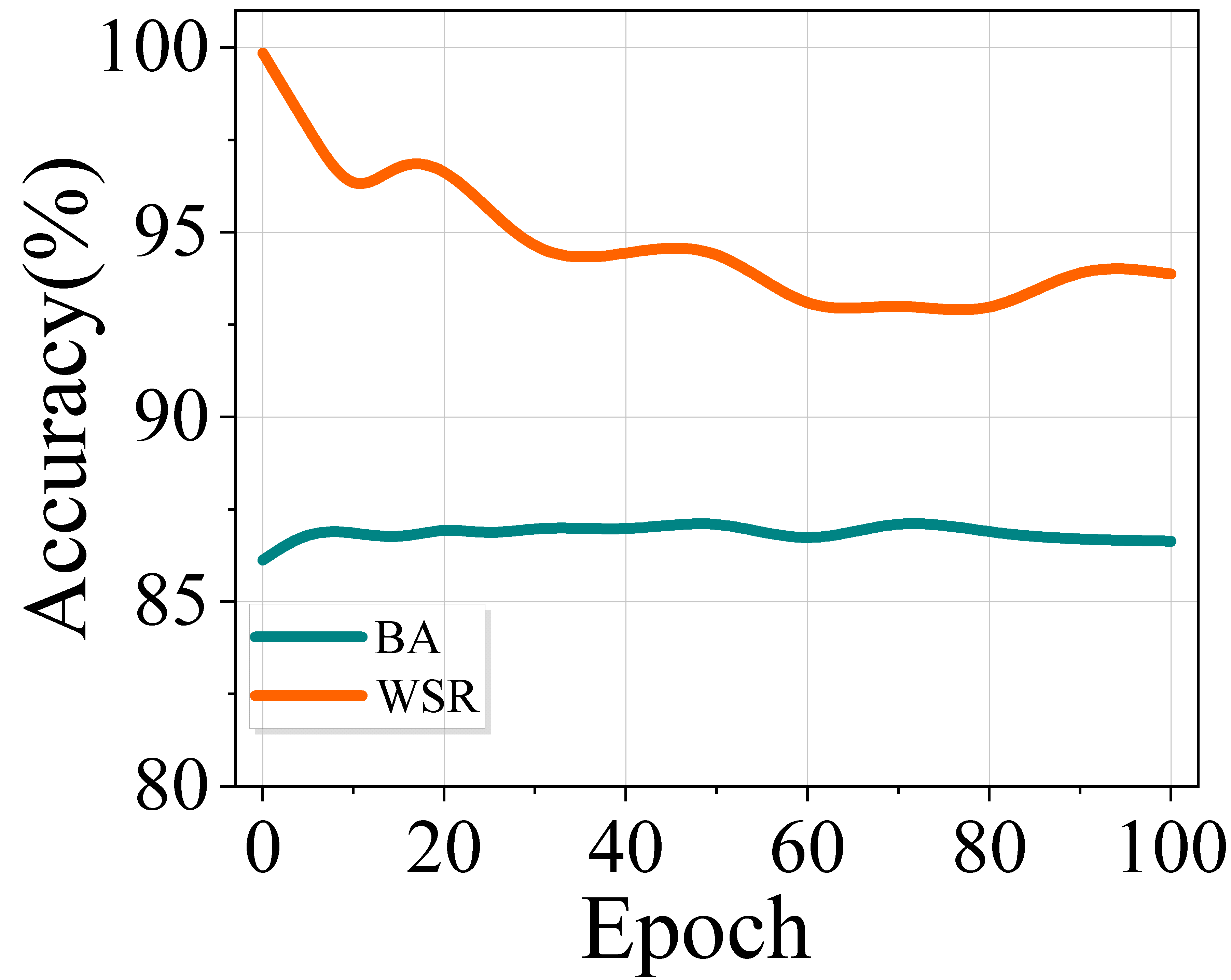}   
    \hspace{-0.5em}  
    \includegraphics[width=0.22\textwidth]{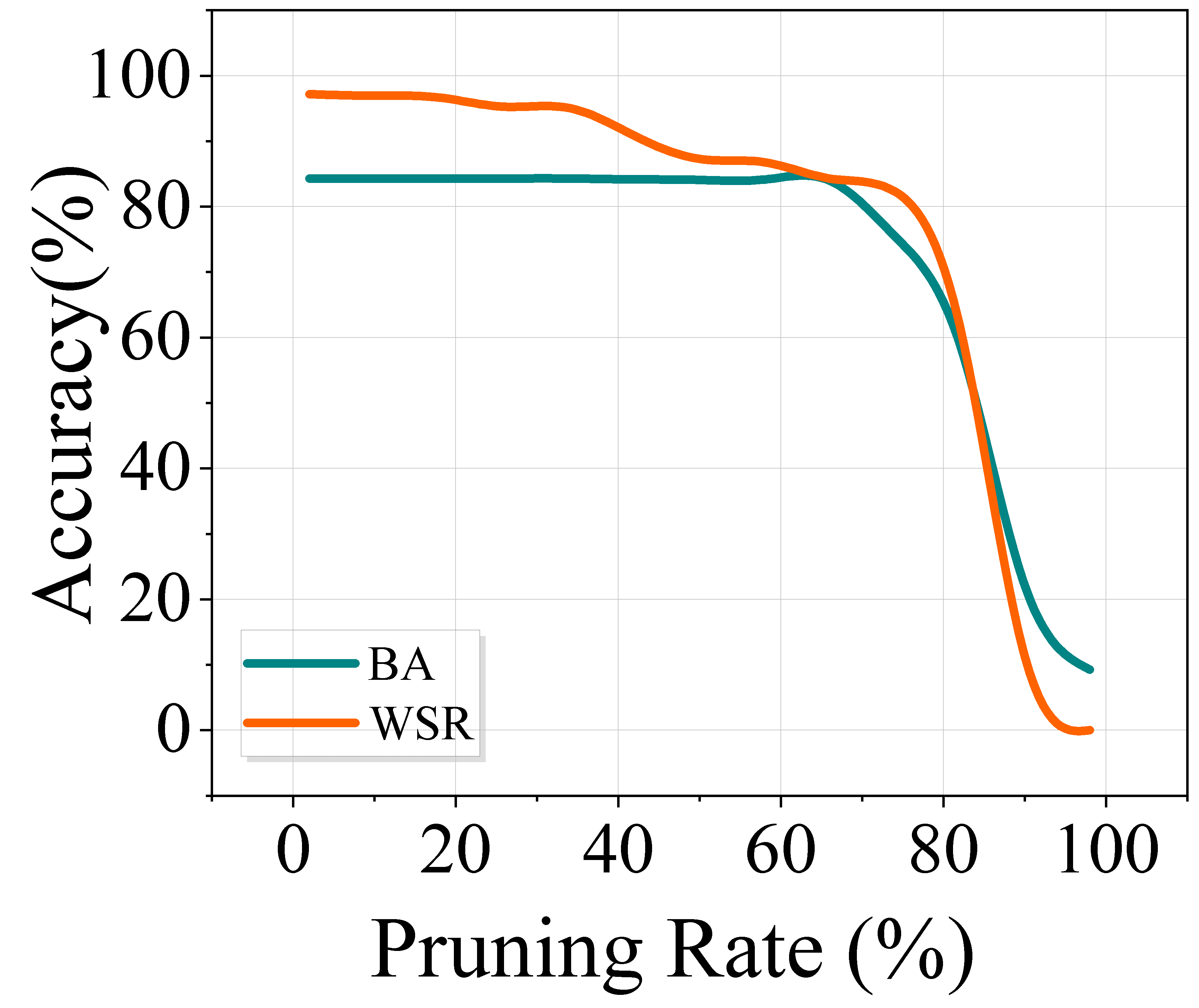}
\caption{Resistance of SSCL-BW to fine-tuning and pruning} 
\label{fig:finetune&pruning}
\vspace{-1.2em}
\end{figure}

\subsection{The Resistance to Potential Watermark-removal Attacks}
When potential malicious users learn that the dataset owner might employ the watermarking approach proposed in this paper for dataset watermarking, they may attempt to launch watermark removal attacks to avoid detection. In this section, we exploit CIFAR-10 and ResNet-18 as an experimental example, to systematically evaluate the resistance of our SSCL-BW against various watermark removal attacks. We discuss two representative and the most widely used watermark removal methods, including fine-tuning \cite{liu2017neural} and model pruning \cite{liu2018fine}.

\noindent \textbf{Resistance to Fine-tuning.} We randomly select 10\% benign samples from the training set to fine-tune the watermarked model. The epoch of fine-tuning is set to $\{ 0, 20, \cdots 100\} $. As shown in Figure \ref{fig:finetune&pruning}, the watermark success rate (WSR) generally decreases with the increase of fine-tuning epochs. However, the WSR is still above $92\%$ when the fine-tuning process is finished, eventually stabilizing. Fine-tuning has only minor effects in reducing WSR. This result indicate that our method is resistant to model fine-tuning.

\vspace{0.3em}
\noindent \textbf{Resistance to Model Pruning.}  We randomly select 10\% benign samples from the original training set to prune the latent representation (\ie, inputs of the fully-connected layers) of the watermarked model. The pruning rate is set to $\{0\%, 2\%, \cdots 98\%\}$. As shown in Figure \ref{fig:finetune&pruning}, the WSR exhibits a gradual decline as the pruning rate increases. However, the benign accuracy (BA) also drops simultaneously. When the pruning rate exceeds 80\%, the BA begins to decline sharply. That means the model’s normal functionality has been compromised. Moreover, when the pruning rate is below 80\%, our method maintains a WSR above $80\%$. This result indicate that our method is resistant to model pruning to some extent.

\subsection{The Model Transferability of SSCL-BW}
In training the generator $G(x;\theta)$, we require a classifier model $f(x;\omega)$ (\ie, surrogate model). In our experiments, we test our method under the same model architecture used to train $G(x;\theta)$. However, compared to the actual model used to generate the watermarked dataset, the suspicious model may have a different structure. In this section, we study the impact of the structural differences between the surrogate model and the suspicious model on the performance of our SSCL-BW.

\noindent \textbf{Settings.}
We select four typical neural networks as benchmark models including ResNet-18 and ResNet-34 \cite{he2016deep}, and VGG-16 and VGG-19 \cite{simonyan2014very}). The experiments adopt a full cross-validation design, where each model is alternately used as the suspicious model and the surrogate model.

\begin{figure}[!t]
\centering
    \vspace{-1em}
        \subfigure[BA]{
        \includegraphics[width=0.22\textwidth]{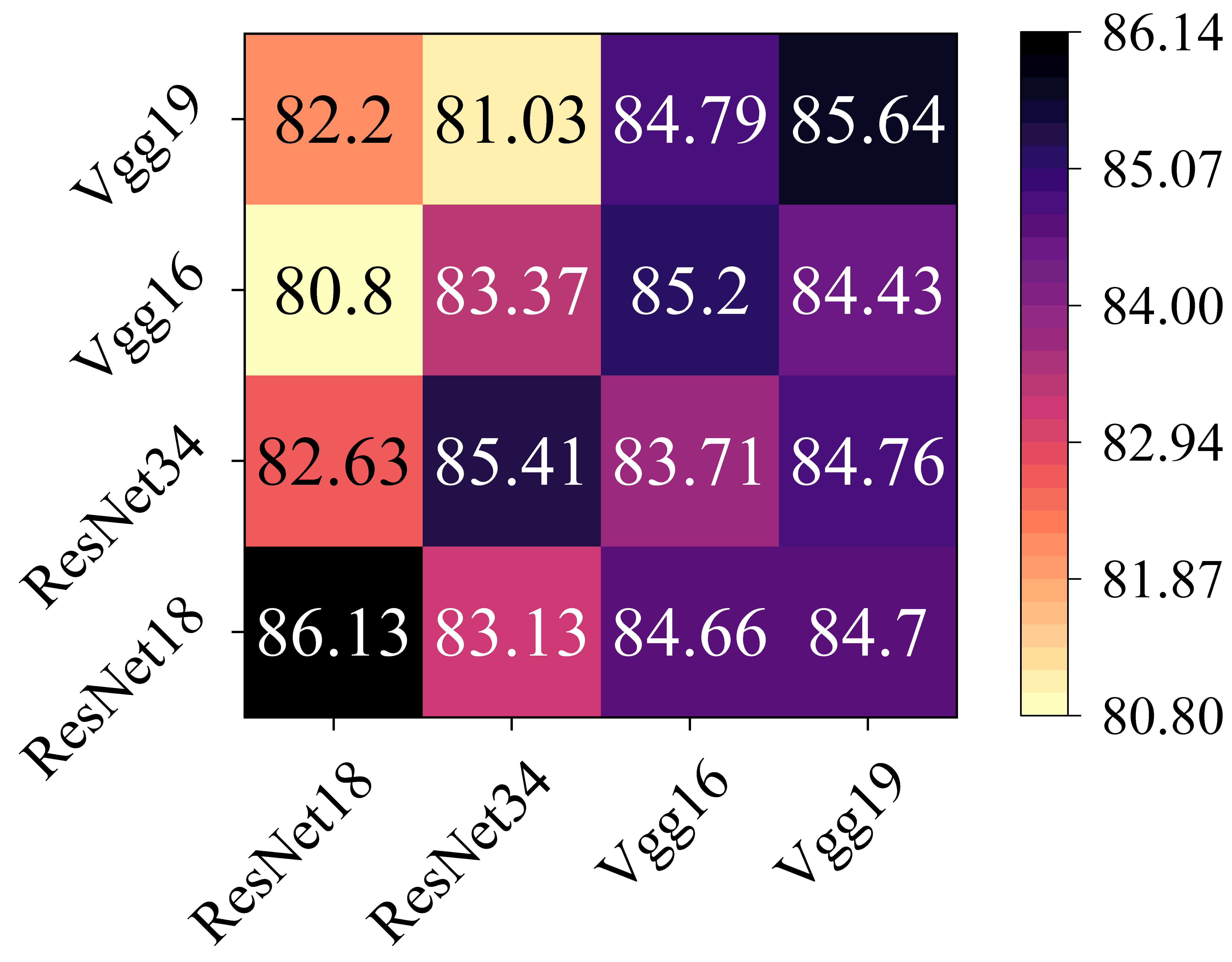}
        }
        \hspace{-0.8em}
        \subfigure[WSR]{
        \includegraphics[width=0.22\textwidth]{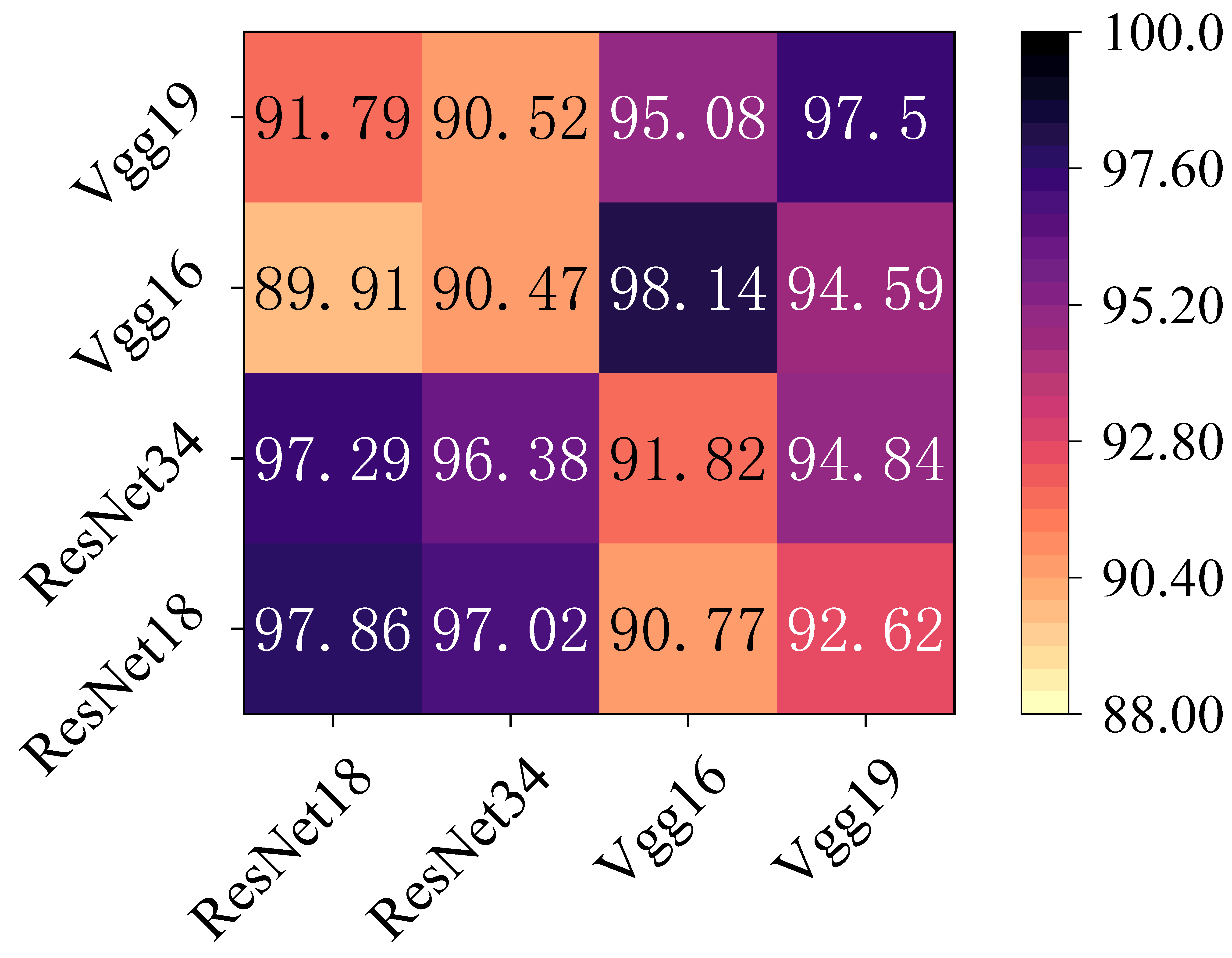}
        }
    \caption{Transferability of SSCL-BW dataset watermarking across different surrogate and training model structures.} 
    \label{fig:change_model}
    \vspace{-1.0em}
    \end{figure}

\noindent \textbf{Results.}
As shown in Figure \ref{fig:change_model}, although the surrogate model and the suspect model have different architectures, our method still maintains high watermark success rate (WSR) and benign accuracy (BA). Our watermark remains highly effective across various scenarios. These results verify the transferability of our SSCL-BW. In other words, the dataset owner does not need to concern the specific models used by the users.

    \section{Conclusion}
    \label{sec:conclusion}
    This paper proposes SSCL-BW, an innovative sample-specific clean-label backdoor watermarking method for dataset ownership verification. Through a U-Net-based watermark generator and a tailored three-component loss function—comprising target sample loss, non-target sample loss, and perceptual similarity loss—this method embeds unique and imperceptible watermarks into individual samples while preserving label consistency, successfully addressing the limitations of existing methods in terms of stealthiness and robustness. Experimental results demonstrate that SSCL-BW significantly outperforms existing clean-label watermarking methods in effectiveness, remains competitive with poison-label approaches, and exhibits strong robustness against common watermark removal attacks, as well as excellent transferability across model architectures. Future work will explore extending this method to cross-modal data and integrating blockchain technology to achieve more transparent and traceable copyright management.
    
    \bibliographystyle{IEEEtran}
    
    \bibliography{reference}
    \end{document}